\documentclass{svjour3} 
\smartqed

\usepackage[cmex10]{amsmath}
\usepackage{amssymb}
\usepackage{wrapfig}
\usepackage{graphicx}

\newcommand{\bma}[1]{\mbox{\boldmath${#1}\/$}}
\newcommand{\Cdot}{\bma{\cdot}}
\newcommand{\Nabla}{\bma{\nabla}}

\begin{document}

\title{Neural network interpolation of the magnetic field for the 
       LISA Pathfinder Diagnostics Subsystem}

\author{Marc~Diaz-Aguilo \and
        Alberto~Lobo \and
        Enrique~Garc\'\i a--Berro
}

\institute{Marc~Diaz--Aguilo and Enrique~Garc\'\i a--Berro  \at
           Departament de F\'\i sica  Aplicada,\\ 
           Universitat Polit\`ecnica  de Catalunya,\\
           c/Esteve Terrades 5, 08860 Castelldefels, Spain\\
           Institut d'Estudis Espacials de Catalunya,\\
           c/Gran Capit\`a 2--4, Edif. Nexus 104,   
           08034 Barcelona, Spain\\
           \email{marc.diaz.aguilo@fa.upc.edu, garcia@fa.upc.edu}
           \and
           Alberto Lobo \at
           Institut de Ci\`encies de l'Espai, CSIC,\\
           Campus UAB, Facultat de Ci\`encies,\\
           Torre C-5,  08193 Bellaterra, Spain\\
           Institut  d'Estudis  Espacials  de Catalunya, \\
           c/Gran Capit\`a 2-4, Edif. Nexus 104,
           08034 Barcelona, Spain\\
           \email{lobo@ice.cat}}

\date{Received: \today / Accepted: \today}

\maketitle

\begin{abstract}
LISA  Pathfinder  is a  science  and  technology  demonstrator of  the
European Space Agency within the  framework of its LISA mission, which
aims to be the  first space-borne gravitational wave observatory.  The
payload of  LISA Pathfinder is the so-called  LISA Technology Package,
which is  designed to measure relative accelerations  between two test
masses in nominal free fall.  Its disturbances are monitored and dealt
by  the diagnostics  subsystem.   This subsystem  consists of  several
modules, and  one of these  is the magnetic diagnostics  system, which
includes a  set of four tri-axial fluxgate  magnetometers, intended to
measure with high precision the magnetic field at the positions of the
test masses.   However, since the  magnetometers are located  far from
the  positions  of  the  test  masses, the  magnetic  field  at  their
positions  must be  interpolated.   It has  been  recently shown  that
because   there   are   not   enough  magnetic   channels,   classical
interpolation  methods fail  to  derive reliable  measurements at  the
positions of  the test masses, while neural  network interpolation can
provide the  required measurements at  the desired accuracy.   In this
paper  we expand  these  studies  and we  assess  the reliability  and
robustness of  the neural network interpolation  scheme for variations
of the locations and possible offsets of the magnetometers, as well as
for changes in environmental conditions.  We find that neural networks
are  robust enough  to derive  accurate measurements  of  the magnetic
field at the positions of the test masses in most circumstances.


\end{abstract}


\section{Introduction}
\label{chap.1}

LISA (Laser  Interferometer Space Antenna)  is a joint  ESA/NASA space
mission aimed  at detecting low frequency gravitational  waves, in the
range between 10$^{-4}$\,Hz and 1\,Hz. LISA will be a constellation of
three spacecraft which occupy the vertexes of an equilateral triangle,
with  sides   of  5  million   kilometers.   The  barycenter   of  the
constellation will revolve  around the Sun in a  quasi circular orbit,
inclined  1$^\circ$ with  respect to  the ecliptic,  and  trailing the
Earth  by   some  20$^\circ$,  about  45   million  kilometres.   Each
spacecraft  harbors  two  proof  masses, carefully  protected  against
external  disturbances such  as solar  radiation pressure  and charged
particles,  which  ensures  they  are  in  nominal  free-fall  in  the
interplanetary  gravitational field.  Gravitational  waves show  up as
differential accelerations  between pairs  of proof masses  in distant
spacecrafts,  and the  working principle  of LISA  is to  measure such
accelerations  using picometer  precision  laser interferometry.   The
interested   reader    is   referred   to   references~\cite{bib:axel}
and~\cite{bib:bill} for more extensive  information, as well as to the
LISA International Science Team (LIST) webpage~\cite{bib:web}.

The  technologies required  for the  LISA  mission are  many and  very
challenging.  This,  coupled with the  fact that some  flight hardware
cannot  be tested  on  ground  since free  fall  conditions cannot  be
maintained  during  periods of  hours,  led  to  set up  a  technology
demonstrator  to   test  critical   LISA  technologies  in   a  flight
environment.   These  technologies  will  be  tested  in  a  precursor
mission, which is called LISA Pathfinder (LPF). This mission is framed
within  the  Scientific Program  of  ESA, and  it  is  expected to  be
launched towards early  2012.  The idea of LPF is  to squeeze one LISA
arm from five million kilometers to 35 centimeters, then determine the
noise of the measurements in  a frequency range which is slightly less
demanding than  that of LISA.   More specifically, the  requirement is
formulated in terms of spectral density of acceleration noise as
\begin{equation}
 S_{\delta a}^{1/2} \leq 3\times 10^{-14}\,\left[1 +
 \left(\frac{\omega/2\pi}{3 {\rm mHz}}\right)^2\right]\ 
 {\rm m\,s}^{-2}\,{\rm Hz}^{-1/2}
 \label{eq.0}
\end{equation}
for        1\,mHz\,$\leq$\,$\omega/2\pi$\,$\leq$\,30\,mHz,       where
$\omega/2\pi$ is the frequency in Hz.

\begin{figure}[!t]
\centering
\includegraphics[width=0.6\columnwidth]{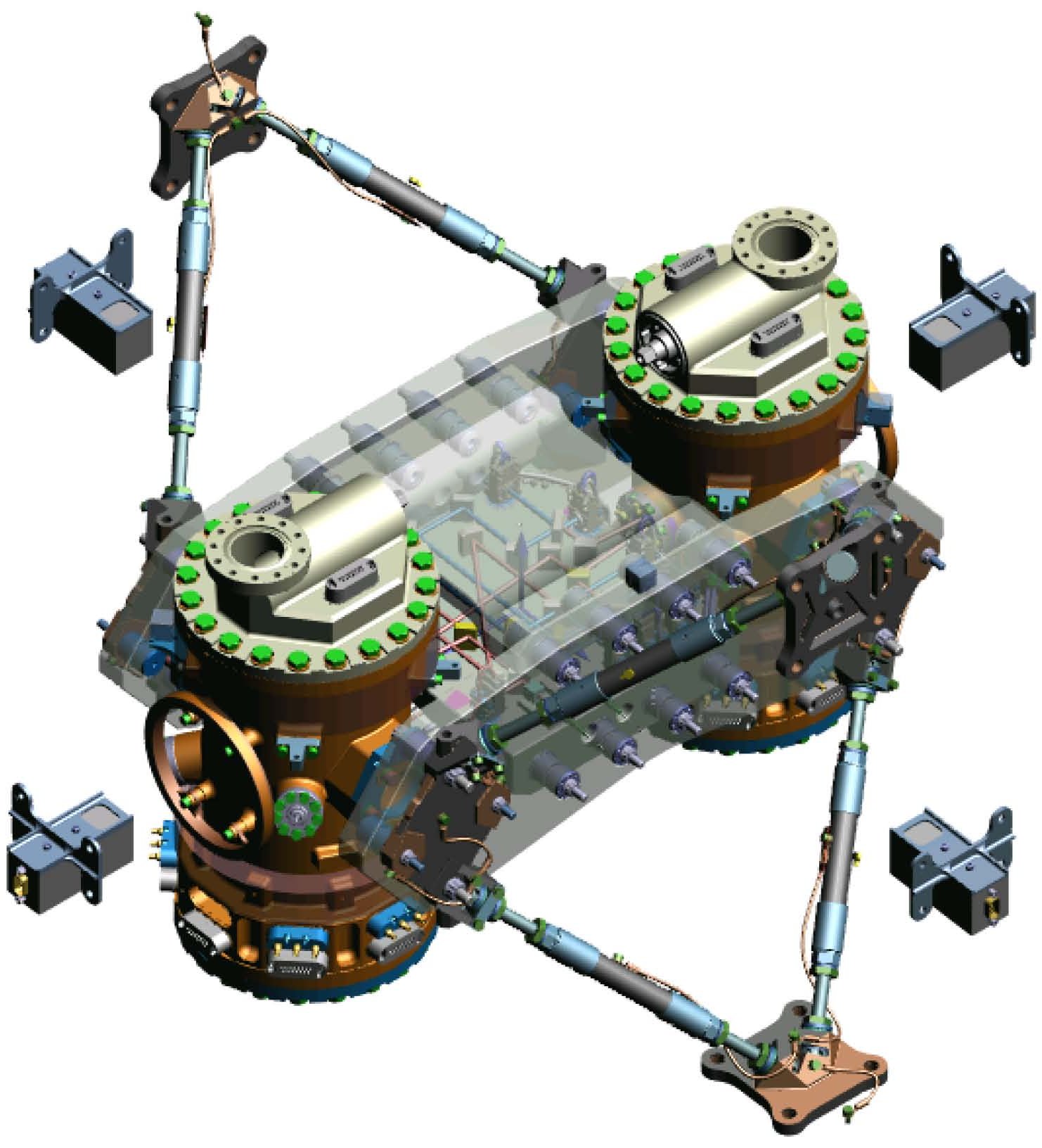}
\caption{A schematic view of the payload of LISA Pathfinder, the LTP.}
\label{fig:LTP}
\end{figure}

The  payload  on board  LPF  is  called  the LISA  Technology  Package
(LTP)~\cite{bib:anza}. Its  main components are  the two gravitational
reference  sensors, which  are  the two  large  vertical cylinders  in
figure~\ref{fig:LTP}. In this  figure it can also be  seen the optical
metrology subsystem,  which provides picometer  precision measurements
of  the relative  position and  acceleration of  the two  test masses,
using precise interferometry.  This  system is located between the two
gravitational reference sensors.  Also visible, and specially relevant
here, are  the four tri-axial magnetometers, which  are represented as
floating boxes. Their actual physical support is the lateral wall of a
(not drawn) larger cylinder which encloses the entire LTP.

Magnetic noise in the LTP is required  to be not more than 40\% of the
total        readout       noise,       i.e.,        $1.2       \times
10^{-14}$\,m\,s$^{-2}$\,Hz$^{-1/2}$     out     of     $3.0     \times
10^{-14}$\,m\,s$^{-2}$\,Hz$^{-1/2}$ in  the measurement bandwidth, see
Eq.~(\ref{eq.0}).    This   noise   appears   because   the   residual
magnetization and  susceptibility of the  test masses couple  with the
surrounding magnetic field, giving rise to a force in each of them
\begin{equation}
 {\bf F} = \left\langle\left[\left({\bf M} +
 \frac{\chi}{\mu_0}\,{\bf B}\right)\Cdot\Nabla\right]{\bf B}\right\rangle V.
 \label{eq.1}
\end{equation}
In this  expression {\bf B}  is the magnetic  field in the  test mass,
{\bf M} is its density  of magnetic moment (magnetization), $V$ is the
volume  of  the test  mass,  $\chi$  is  its magnetic  susceptibility,
$\mu_0$  is the  vacuum magnetic  constant  ($4\pi\times 10^{-7}$\,m\,
kg\,  s$^{-2}$\, A$^{-2}$),  and $\langle\cdots\rangle$  indicates the
volume  average  of  the  enclosed  quantity.  As  clearly  seen  from
Eq. (\ref{eq.1}), there are two  sources of magnetic noise.  The first
one is due to the fluctuations  of the magnetic field and its gradient
in  the regions  occupied  by the  test  masses \cite{bib:LISA}.   The
second  one comes from  the susceptibility  of the  test mass  and the
magnetic  remanence   fluctuations~\cite{bib:ntcs}.   This  additional
noise  is  expected to  be  much less  important,  and  it is  usually
disregarded.

Clearly,  a  quantitative assessment  of  magnetic  noise  in the  LTP
requires  real-time   monitoring  of   the  magnetic  field   and  its
gradient~\cite{bib:trento}.   This   is  the  ultimate   goal  of  the
tri-axial  fluxgate  magnetometers~\cite{bib:DDS_LTP}.  These  devices
have  a  high  permeability  magnetic  core,  which  drives  a  design
constraint  to keep  them somewhat  far from  the test  masses.  Thus,
their readouts  do not  provide a direct  measurement of  the magnetic
field at  the position  of the test  masses, complicating the  task of
inferring the field at  their position, and forcing the implementation
of  an interpolation  method to  overcome this  shortcoming.  However,
such  interpolation process faces  serious difficulties.   Indeed, the
size of  the interpolation  region, that is,  the interior of  the LTP
Core Assembly (LCA), is too large for a linear interpolation scheme to
be reliable.   Additionally, the number of  magnetometer channels does
not provide sufficient  data to go beyond a  poor linear approximation
\cite{bib:myPaper}.  However,  the structure of the  magnetic field is
rather complex, as  the sources of magnetic field  are essentially the
electronic components inside the  spacecraft. The number of identified
sources is  about 50,  and they behave  as magnetic dipoles,  the only
exception  being the  solar panel,  which  is best  approximated by  a
quadrupole.   The  positions  of  the  sources  are  dictated  by  the
architecture  of the satellite,  which defines  the exact  position of
each  electronic  subsystem.  Fortunately,  there  are  no sources  of
magnetic field inside the LCA, all being placed within the spacecraft,
but outside the  LCA walls.  Adequate processing of  all the available
information  shows that  the  magnetic field  is  smaller towards  the
center of the  spacecraft (where the test masses  are located) than it
is in its periphery (where the magnetometers take measurements).

It has been recently  shown \cite{bib:myPaper} that since the standard
interpolation  scheme, which is  based in  multipole expansion  of the
magnetic field  inside the LCA  volume, does not go  beyond quadrupole
order,  its  performance in  estimating  the  magnetic  field and  its
gradients is  very poor. On  the contrary, artificial  neural networks
have been shown to be  a reliable alternative to estimate the required
field and  gradient values at the  positions of the  test masses.  The
reasons for this are  multiple.  Firstly, the multipole expansion only
takes  into account  the readings  of the  magnetometers,  whereas the
artificial neural network  also uses the actual value  of the magnetic
field at the position of the test masses to train the network. This is
a  crucial  issue  since  the  interpolation  algorithm  is  fed  with
additional information.  Secondly,  the classical interpolation method
seeks  for a  global solution  of the  magnetic field.   That  is, the
multipole expansion models the magnetic field inside the entire volume
of  the  LCA.   Clearly,  since  the  available  information  for  the
multipole  expansion is  rather  limited, the  quality  of the  global
solution  is very  poor.   In sharp  contrast,  the artificial  neural
network first finds and then uses the correlation between the magnetic
field at  the positions  of the magnetometers  and the test  masses to
obtain  reliable  values  of  the  magnetic  field  for  any  magnetic
configuration.   As a matter  of fact,  the artificial  neural network
performs  a  point-to-point  interpolation  and  it is  not  aimed  at
reproducing the highly non-linear magnetic field well at any arbitrary
position  within the volume  of the  LCA.  Finally,  artificial neural
networks  are   trained  using  a  large  number   of  magnetic  field
realizations,  thus  the interpolating  algorithm  uses a  statiscally
elaborated  information.  In this  sense, it  is important  to realize
that artificial  neural networks  have been shown  to be a  robust and
easily  implementable technique  among  numerous statistical  modeling
tools  \cite{bib:kecman}.  On  the contrary,  the  multipole expansion
does  not  use statistical  information.   Once  the  readings of  the
magnetometers  are known,  the theoretical  solution for  the magnetic
field  within  the  entire  volume  of  the LCA  is  determined  in  a
straightforward way.

Nevertheless,  an   in  depth  study   of  how  the  results   of  the
interpolation procedure depend on  the specific characteristics of the
neural  network  remains  to  be  done. It  also  remains  to  further
investigate  why the  neural network  --- which  uses  lineal transfer
functions --- obtains such good results interpolating the value of the
magnetic field  at the  positions of the  test masses, which  are well
inside a deep  well of magnetic field.  Finally,  an assessment of the
robustness of the neural network  interpolating scheme in front of the
unavoidable errors in the positions  of the magnetometers, or in front
of  low-frequency variations  of  the magnetic  environment and,  more
importantly, in front of offsets  in the readings of the magnetometers
still is needed. These are precisely the goals of this paper.

The paper  is organized as follows.  In  Sect.~\ref{chap.2} we discuss
the  appropriateness of  our neural  network approach  to  measure the
magnetic field and its gradients  at the positions of the test masses,
and  we  discuss which  are  the  accuracies  obtained when  different
architectures  of   the  neural  network  are   adopted.   It  follows
Sect.~\ref{chap.3}, where we discuss how the unavoidable errors in the
onground measurements  of the magnetic dipoles of  each electronic box
affect   the  performance   of   the  adopted   neural  network.    In
Sect.~\ref{chap.4} we evaluate the  expected errors in the estimate of
the magnetic field  and its gradients due to a  possible offset in the
readings  of the  magnetometers  due to  launch  stresses, whereas  in
Sect.~\ref{chap.5}  we  study  how  the mechanical  precision  of  the
positions of the tri-axial  magnetometers and their spatial resolution
affect  the determination  of the  magnetic field  and  its gradients.
Sect.~\ref{chap.6} is devoted to  assess the reliability of our neural
network approach  in front of  a slowly varying  magnetic environment.
Finally,  in Sect.~\ref{chap.7}  we  summarize our  main findings,  we
discuss the significance of our results and we draw our conclusions.


\section{The neural network architecture}
\label{chap.2}

Although   neural  networks   have  been   used  in   different  space
applications  \cite{bib:scatterometer,bib:satellite}, to  the  best of
our  knowledge this  is the  first application  of neural  networks to
analyze  inflight  outputs in  space  missions.   Hence, studying  the
robustness of the neural network architecture proposed to estimate the
magnetic field inside the LCA is a mandatory task.

\subsection{The fiducial neural network architecture}

\begin{figure}[!t]
\centering
\includegraphics[width=0.6\columnwidth]{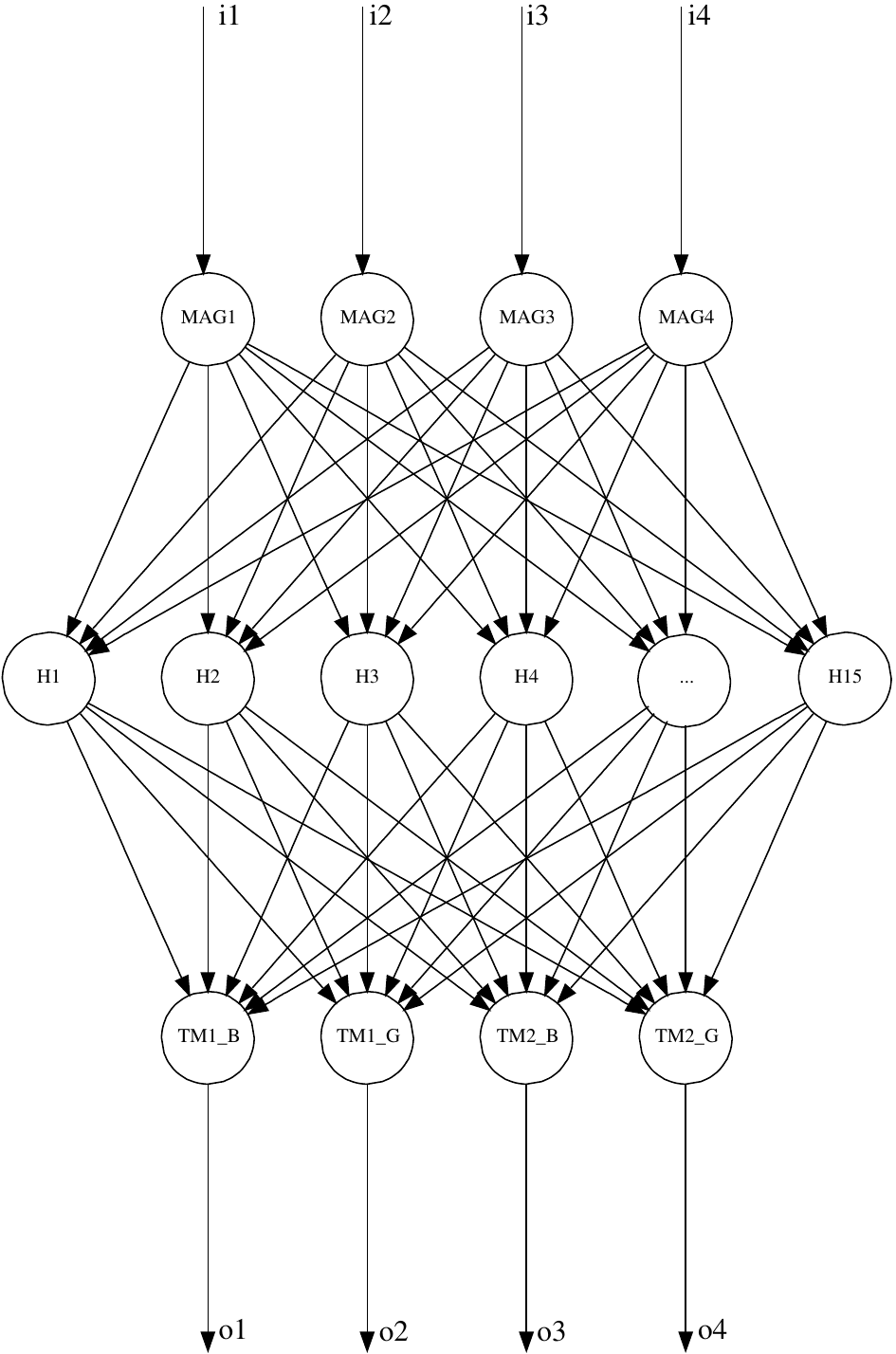}
\caption{The  fiducial feed-forward  neural network  architecture. The
         readings  of  the  magnetometers  are the  system  inputs  (4
         magnetometers, each one with 3 data channels). The outputs of
         the system are the  magnetic field and gradient components at
         the positions  of the test  masses (3 field components  and 5
         gradient  estimates for  each  test mass).  For  the sake  of
         simplicity,  all the  field and  gradient channels  have been
         grouped into  a single neuron. Moreover, not  all the neurons
         in the hidden layer are shown in this figure.}
\label{fig:neuralarquitecture}
\end{figure}

Fig.  \ref{fig:neuralarquitecture}  shows a simplified  version of the
fiducial  architecture of  our neural  network.  As  can be  seen, the
number of inputs  is twelve --- one for  each magnetometer readout ---
corresponding  to  the  four  tri-axial magnetometers  placed  in  the
spacecraft.  These  readings are  the only valuable  information which
can be  used to estimate  the magnetic field  at the positions  of the
test masses, and constitute the input layer of the neural network.  On
the other hand,  to estimate the magnetic field  three outputs will be
required --- corresponding to the three field components per test mass
--- whereas to estimate the  gradient only five additional outputs are
needed. This  is because  the magnetic field  has zero  divergence and
zero  rotational.  Thus,  the gradient  matrix  $\partial B_i/\partial
x_j$ is a traceless symmetric matrix,  and therefore only 5 out of its
9 components are  independent.  These outputs are the  output layer of
the  neural network.   In  addition to  the  two previously  described
layers, there  is only one  intermediate layer, which  constitutes the
hidden  layer.  This   layer  is  made  of  15   neurons.  Using  this
architecture  for  the neural  network  the  magnetic field  estimates
typically  have  standard  deviations  on  the  order  of  $\sim  2\%$
\cite{bib:myPaper}, a  value to  which we compare  the results  of our
analysis.

\subsection{Training and testing}
\label{sec:model}

Training and testing data sets  were simulated using the most complete
and up-to-date information about the magnetic configuration within the
spacecraft. The  complete magnetic configuration of  the satellite has
not been measured yet, because  some units have not been delivered yet
to  the prime contractor.   Nevertheless, the  exact position  of each
unit in the  spacecraft reference frame is known.   On the other hand,
the magnetic moments used in our simulations are those reported by the
constructors  of  each subsystem.   Unfortunately,  this  data is  not
available yet for  all units, and moreover although  the moduli of the
dipoles  are known  for all  the subsystems  their directions  are not
known yet for most of  the units.  The three-dimensional values of the
magnetic dipoles of each unit will be accurately measured in the final
testing campaign to be  performed on each subsystem before assembling.
This campaign is expected to  be performed on the assembled spacecraft
during 2011.  The training and  validation of the neural network using
the measured  values of  the magnetic dipoles  will be done  after the
campaign  but the  specific details  of the  processing  algorithm are
expected to remain unchanged.  Moreover, the magnetic field inside the
LCA  is   expected  to   vary  substantially  between   the  different
operational modes.   Accordingly, since the  magnetic configuration of
the  spacecraft  may  have  different  characteristics  for  different
operational modes, it is foreseen that a different neural network will
be trained for each of these configurations.

Given  the unknown orientations  of the  magnetic dipoles  we generate
several magnetic configurations assigning randomly the orientations of
the  46 dipoles.   An  example  scenario is  thus  characterized by  a
selection  of the  46 dipoles  with random  orientations.   The random
character  of the  procedure may  seem unrealistic,  since  the actual
satellite  configuration is  not  random.  In  this context,  however,
randomness is an  efficient way of mimicking our  lack of knowledge of
all the directions of the  sources of magnetic field. Nevertheless, it
is worth  mentioning that  the produced sets  are consistent  with our
expectations and  the mission requirements \cite{bib:LISA,bib:DDS_LTP}
since at  the positions of the  test masses we  obtain magnetic fields
$\sim  300$ nT, while  the readings  at the  magnetometers are  of the
order  of 4  to  10 $\mu$T.   With  this approach  the magnetic  field
generated by  the dipole distribution at  a generic point  {\bf x} and
time $t\/$ is therefore given by
    \begin{equation}
    {\bf B}({\bf x},t) = \frac{\mu_0}{4\pi}\;\sum_{a=1}^{n}
    \frac{3\left[{\bf m}_a(t)\Cdot {\bf n}_a\right]\,{\bf n}_a - 
    {\bf m}_a(t)}{\left|{\bf x}-{\bf x}_a\right|^3}
    \label{eq.15}
    \end{equation}
where  {\bf n}$_a$\,=\,  $\left({\bf  x}-{\bf x}_a\right)$/$\left|{\bf
x}-{\bf x}_a\right|$ are unit vectors connecting the the $a$-th dipole
{\bf m}$_a$ with  the field point {\bf x}, and $n\/$  is the number of
dipoles.   In order  to simulate  realistic magnetic  environments, we
compute the magnetic  field at the positions of  the magnetometers and
at the  positions of  the test masses  using Eq.   (\ref{eq.15}).  The
positions of  the test  masses and of  the magnetometers are  shown in
table  \ref{tab:positions}.  Two different  batches of  $10^3$ samples
are generated.  The first  batch was  used as the  training set  for a
neural     network      with     the     architecture      of     Fig.
\ref{fig:neuralarquitecture}.   This batch  consists in  12  inputs (3
inputs  for  each  of  the  4 vector  magnetometers)  and  16  outputs
representing the  field information  at the position  of the  two test
masses (3 field plus 5 gradient components per test mass).  The second
batch has  been used for validation  to assess the  performance of the
neural network.

\begin{table}[t]
\caption{Positions   of  the   test  masses   and  positions   of  the
         magnetometers referred  to a  coordinate system fixed  to the
         spacecraft.}
\label{tab:positions}
\begin{center}
\begin{tabular}{l r r r}
 \hline\noalign{\smallskip}
 Test masses &  $x$ [m] & $y$ [m] & $z$ [m]  \\
 \noalign{\smallskip}\hline\noalign{\smallskip}
 1 & $-$0.1880 & 0 & 0.4784 \\
 2 &    0.1880 & 0 & 0.4784 \\
 \noalign{\smallskip}\hline\noalign{\smallskip}
Magnetometers &  $x$ [m] & $y$ [m] & $z$ [m]  \\
 \noalign{\smallskip}\hline\noalign{\smallskip}
 1 & $-$0.0758 & 0.3694    & 0.4784 \\
 2 & $-$0.3765 & 0         & 0.4784 \\
 3 &    0.0758 & $-$0.3694 & 0.4784 \\
 4 &    0.3765 & 0         & 0.4784 \\
\noalign{\smallskip}\hline
\end{tabular}
\end{center}
\end{table}

\subsection{Varying the number of neurons}

\begin{figure}[!t]
\centering
\includegraphics[width=0.9\textwidth]{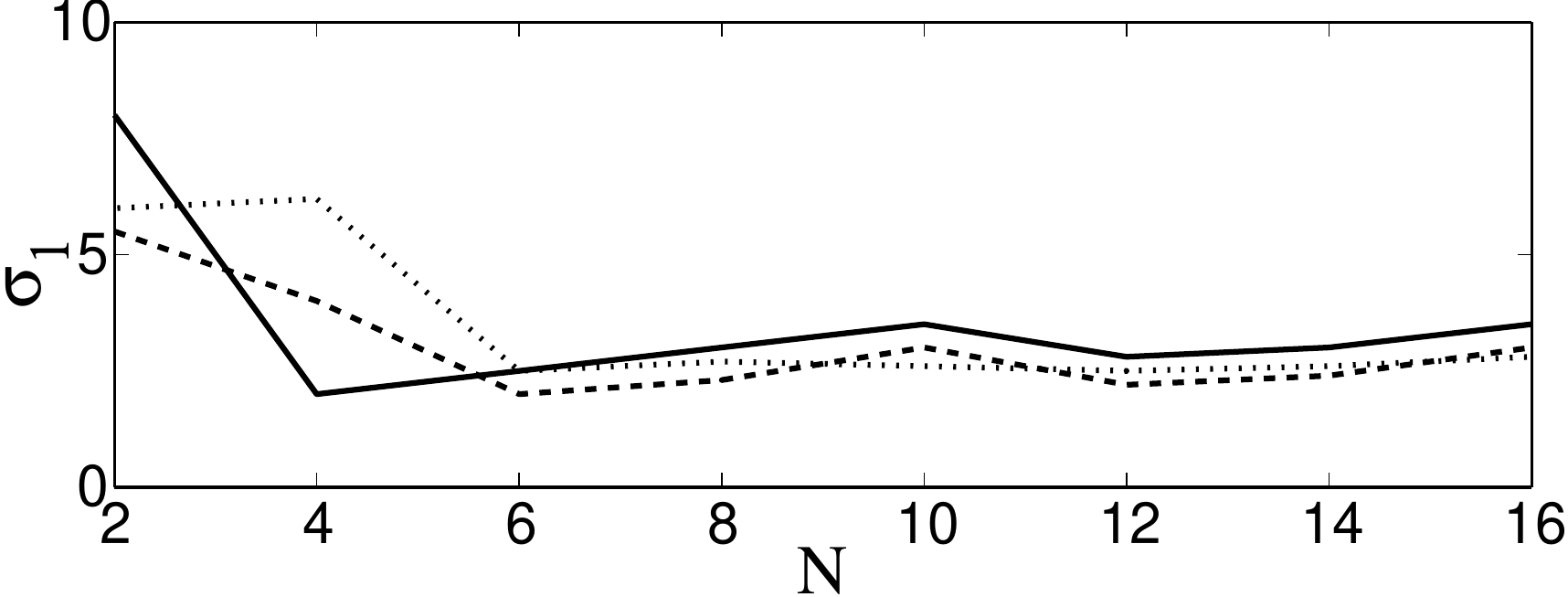}
\includegraphics[width=0.9\textwidth]{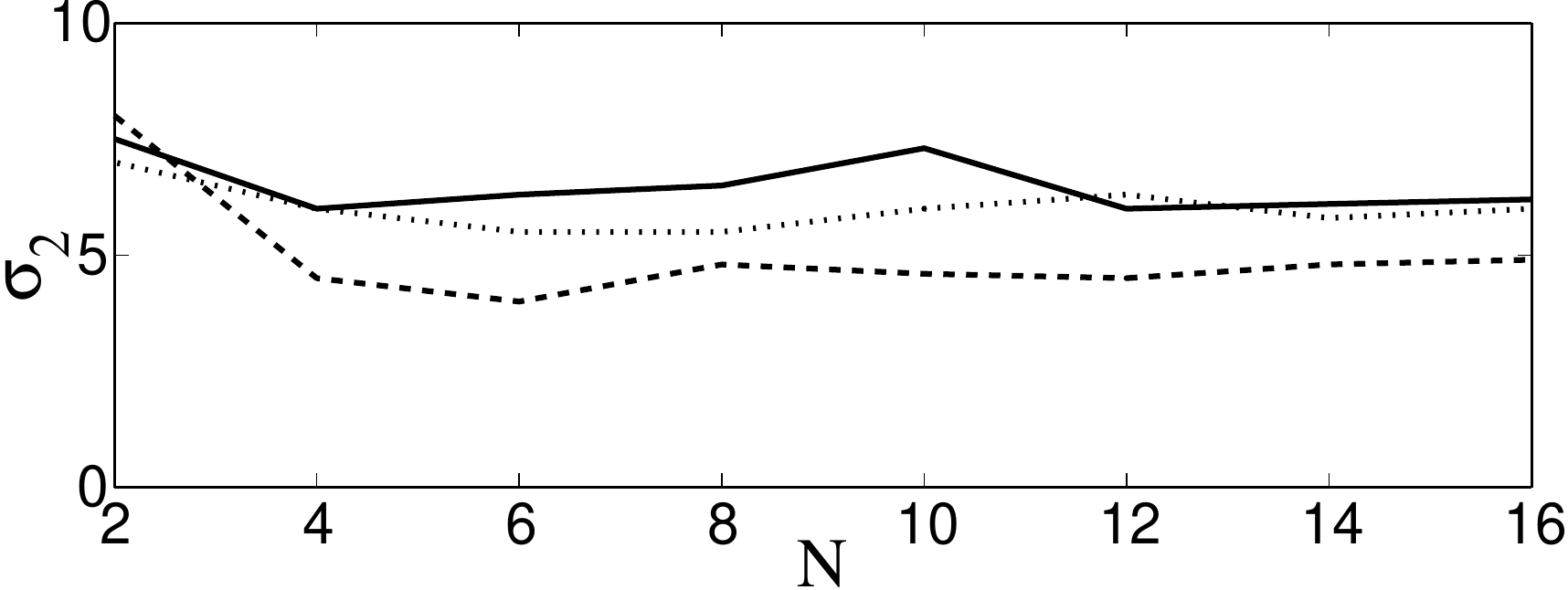}
\caption{Quality of the  estimate of the magnetic field  as a function
         of the  number of neurons  in the hidden layer.   The maximum
         interpolation  error  remains   almost  constant  for  neural
         networks larger  than $\sim 15$ neurons in  the hidden layer.
         The solid line corresponds to the $B_x$-component, the dotted
         line  to the  $B_y$-component  and, the  dashed  line to  the
         $B_z$-component.}
\label{fig:numberNeuronsField}
\end{figure}

Assesing  the correct  choice of  the number  of neurons  of  a neural
network is not a simple task.   When the neural network is composed by
only one hidden layer, the input layer contains as many inputs-neurons
as  the   information  we   provide  to  the   network  and   as  many
output-neurons  as  the target  information  we  want to  reconstruct.
Nevertheless, as far  as the number of neurons of  the hidden layer is
concerned, it is not guaranteed  that the architecture of the selected
neural network  is optimal  nor there is  an algorithm in  the current
literature     to     determine      the     optimal     number     of
neurons~\cite{bib:neuralPrunning,bib:size}.  Normally,  to obtain good
results, the  smallest system obtained after prunning  that is capable
to fit the data should be used.  Unfortunately, it is not obvious what
size is  best.  A system  with a small  number of neurons will  not be
able to learn from the data,  while one with a large number of neurons
may learn very slowly and, moreover,  it will be very sensitive to the
initial conditions and learning paramaters. Additionally, it should be
taken into account that one  of the biggest problems of large networks
for some  specific problems is  the fact that  in the early  stages of
training, the error  on both the training and  tests tends to decrease
with  time  as  the  network  generalizes  for  the  examples  to  the
underlying function. However, at some  point, the error on the testing
set  reaches a minimum  and begins  to increase  again as  the network
starts  to adapt  to artifacts  and specific  details in  the training
data,  while  the   training  error  asymptotically  decreases.   This
problem,  known  as  overfitting,  occurs  more  frequently  in  large
networks  due  to  the  excessive  number of  degrees  of  freedom  in
comparison to the training set \cite{bib:overfitting}.  To avoid this,
we  have  used the  early  stopping  technique,  which overcomes  this
shortcoming.

In the  early stopping  technique the available  data is  divided into
three subsets \cite{bib:kecman}. The first subset is the training set,
which  is used  for computing  the gradient  and updating  the network
weights and  biases.  The  second subset is  the validation  set.  The
error on the validation set  is monitored during the training process.
The validation  error normally decreases  during the initial  phase of
training, as  does the training  set error. However, when  the network
begins to overfit the data,  the error on the validation set typically
begins to  rise. When the  validation error increases for  a specified
number of  iterations, the  training is stopped,  and the  weights and
biases  at the minimum  of the  validation error  are returned  to the
values  obtained at  the  minimum.  All  these precautionary  measures
avoid overfitting.   Therefore, the analysis of the  number of neurons
needed for the hidden layer can be made analyzing the evolution of the
estimation  error  on  the  testing  set  as  the  number  of  neurons
increases.   The   results  of  such  an  analysis   are  depicted  in
figure~\ref{fig:numberNeuronsField},   which    shows   the   standard
deviation of  the estimate for  both test mass 1,  $\sigma_{\rm {1}}$,
and test  masss 2, $\sigma_{\rm {2}}$  as a function of  the number of
neurons in the hidden layer, $N$.

\begin{table}[!t]
\renewcommand{\arraystretch}{1.3}
\caption{Quality of the estimate for the most common neuron activation
         functions.}
\label{tab:typeNeuron}
\centering
\begin{tabular}{lcccccc}
\hline\noalign{\smallskip}
                           &
\multicolumn{3}{c}{TM$_1$} &
\multicolumn{3}{c}{TM$_2$} \\
\cline{2-7}
 Function &  $\sigma_x$ & $\sigma_y$ & $\sigma_z$ &
             $\sigma_x$ & $\sigma_y$ & $\sigma_z$ \\
\noalign{\smallskip}\hline\noalign{\smallskip}
 Tangent sigmoid     & 4.1 & 3.8 & 2.5 &  5.9 & 5.2 & 4.5 \\
 Linear              & 3.8 & 3.5 & 2.3 &  5.7 & 5.4 & 4.2 \\
 Logarithmic sigmoid & 4.2 & 3.8 & 2.5 &  6.2 & 5.1 & 4.5 \\
 Radial base         & 4.2 & 4.3 & 3.9 &  6.3 & 6.0 & 4.8 \\
 Step                & 7.5 & 7.6 & 4.9 & 12.3 & 8.2 & 7.9 \\
\noalign{\smallskip}\hline
\end{tabular}
\end{table}

As can  be seen in  this figure, when  a reduced number of  neurons is
used the model cannot  accurately estimate the underlying function due
to the lack  of tunnable parameters.  As the number  of neurons in the
hidden layer is increased, the  neural network performs better and for
a number of  neurons larger than 15 the error  is not further reduced.
Consequently,  we  conclude that  for  this  specific application  the
adequate number  of neurons for the  hidden layer lies  between 10 and
15.   This choice  ensures a  network large  enough to  be  capable of
estimating the  underlying relationship  and not excessively  large to
consume excessive training time, learn  slowly and be dependent on the
learning algorithm.   We have also checked that  increasing the number
of hidden  layers of the  neural network does  not result in  a better
performance  of  the interpolating  algorithm,  but  for  the sake  of
conciseness we  do not show  the results here.   All in all,  it seems
that our fiducial architecture seems to work best.

\subsection{Changing the type of neuron}

Most   of   the   feed-forward   networks   are   trained   with   the
back-propagation algorithm and gradient descent techniques are used to
minimize some specific  cost function, and this has  been the case for
the  training algorithm  used here.   This means  that  all activation
functions  within the  network must  be differentiable  to be  able to
compute the  network gradient for  each learning step.   Normally, the
most commonly  used type of functions  are the tangent  sigmoid or the
logarithmic sigmoid~\cite{bib:kecman}, which  can model any non-linear
function   if  properly  trained~\cite{bib:dreyfus},   whereas  linear
functions   are  usually   employed  for   linear  models   with  high
dimensionality.

We have  studied several possibilities  and the results are  listed in
table~\ref{tab:typeNeuron}, where  we show for the  different types of
neurons the  standard deviations of the  probability density functions
of the  estimates of  the magnetic field  for both  test mass 1  and 2
(TM$_1$ and TM$_2$, respectively).  In our case, and as borne out from
table~\ref{tab:typeNeuron},  the  linear  function together  with  the
tangent  sigmoid and the  logarithmic sigmoid  are the  most efficient
choices, while the performance of the radial base function is slightly
worse.  Finally,  the step function (the popular  perceptron) does not
yield good results because it  is specifically designed to be used for
classification problems.   Specifically, the linear neuron  is the one
for which we  obtain the best results. This  could be surprising given
that our problem  is highly non-linear.  The reason  is that for every
magnetic  configuration  there  exists   a  large  and  fairly  stable
difference between the value of  the magnetic field at the location of
the magnetometers  (all of  the components of  the magnetic  field are
$\sim 10$  $\mu$T) and the  field at the  position of the  test masses
(all the  components are on  the 100 nT  level). For this  reason, the
weigths  of  the network  happen  to  be  the most  relevant  modeling
factors.  That is, the  point-to-point interpolation can be understood
in  the linear  case as  a simple  weighted sum  of  the magnetometers
measurements.   Accordingly,  because   of  its  simplicity  and  good
results,  we  use  the  linear  function  as the  basic  unit  in  our
regression  study. It  is  worth  noting at  this  point that  similar
results could  be obtained  using a high-dimensionality  least squares
analysis,  but in  our specific  case we  have found  matrix inversion
problems because some  magnetometer channels present highly correlated
signals.

\subsection{Underlying structures}

We   have  already   shown   that  our   neural   network  is   highly
reliable. Thus,  it is normal to  ask ourselves which  is the ultimate
reason of  this behavior. The answer  to this question  is that during
the training  process, the neural  network eventually learns  that the
magnetic  field at  the  positions  of the  test  masses is  generally
smaller than the magnetometers readouts --- with occasional exceptions
due to the rich and complex  profile structure of the field inside the
LCA.   Moreover, the  neural network  is  able to  learn an  inference
procedure  which is  actually quite  efficient.  To  better understand
this, we found instructive to look into relationships between the data
read  by the  magnetometers and  the estimates  of the  magnetic field
generated by  the neural network.   We chose to  calculate correlation
coefficients  between  input and  output  data,  and  the results  are
displayed  in  table~\ref{table:input-ouput}.   The  test  masses  are
labeled  as   TM$_1$  and   TM$_2$,  respectively,  whilst   the  four
magnetometers are listed as M$_i$, $i=1,\ldots,4$.

\begin{table}[!t]
\caption{Input-output relationship learned by the network.}
\label{table:input-ouput}
\centering
\begin{tabular}{crrr}
\hline\noalign{\smallskip}
Output & $B_x$ & $B_y$ & $B_z$\\
\noalign{\smallskip}\hline\noalign{\smallskip}
$B_x$ TM$_1$ & & & \\
$\quad$ M$_1$ 	&   0.2177  & $-0.1060$ &   0.0134  \\
$\quad$ M$_2$  	&   0.2581  & $-0.0185$ &   0.1564  \\
$\quad$ M$_3$  	&   0.3754  &   0.0985  &   0.0054  \\
$\quad$ M$_4$  	&   0.9340  &   0.1528  & $-0.0501$ \\
\noalign{\smallskip}\hline\noalign{\smallskip}
$B_y$ TM$_1$ & & & \\
$\quad$ M$_1$   & $-0.0197$ &   0.3556  & $-0.0682$ \\
$\quad$ M$_2$   &   0.0031  &   0.2240  &   0.0601  \\
$\quad$ M$_3$   & $-0.0782$ &   0.4249  &   0.1217  \\
$\quad$ M$_4$   &   0.0668  &   0.9035  &   0.0102  \\
\noalign{\smallskip}\hline\noalign{\smallskip}
$B_z$ TM$_1$ & & & \\
$\quad$ M$_1$   & $-0.0772$ & $-0.0635$ &   0.3090  \\
$\quad$ M$_2$   & $-0.1343$ &   0.0083  &   0.3377  \\
$\quad$ M$_3$   & $-0.0180$ & $-0.1027$ &   0.5002  \\
$\quad$ M$_4$   &   0.0493  &   0.0615  &   0.9041  \\
\noalign{\smallskip}\hline\noalign{\smallskip}
$B_x$ TM$_2$ & & & \\
$\quad$ M$_1$   &   0.3506  &   0.1862  &   0.0840  \\
$\quad$ M$_2$   &   0.9081  & $-0.2830$ &   0.3782  \\
$\quad$ M$_3$   &   0.1230  & $-0.2398$ & $-0.0613$ \\
$\quad$ M$_4$   &   0.2502  &   0.0184  & $-0.0480$ \\
\noalign{\smallskip}\hline\noalign{\smallskip}
$B_y$ TM$_2$ & & & \\
$\quad$ M$_1$   & $-0.3662$ &   0.3877  & $-0.0211$ \\
$\quad$ M$_2$   &   0.0184  &   0.8398  & $-0.1200$ \\
$\quad$ M$_3$   &   0.3722  &   0.2400  &   0.0927  \\
$\quad$ M$_4$   & $-0.0040$ &   0.2379  & $-0.0026$ \\
\noalign{\smallskip}\hline\noalign{\smallskip}
$B_z$ TM$_2$ & & & \\
$\quad$ M$_1$   &   0.1217  &   0.0267  &   0.4111  \\
$\quad$ M$_2$   &   0.0144  & $-0.1222$ &   0.8740  \\
$\quad$ M$_3$   &   0.0333  &   0.0233  &   0.5054  \\
$\quad$ M$_4$   &   0.0310  &   0.0141  &   0.2685  \\
\noalign{\smallskip}\hline
\end{tabular}
\end{table}

The following  major features can be easily  identified. Firstly, each
component of  the field is basically estimated  from the magnetometers
reading of the  same component. For example, the  interpolation of the
$B_x\/$-component  in   test  mass  1  is  mostly   dependent  on  the
$B_x\/$-readings of the  magnetometers.  Secondly, the measurements of
the  magnetometers  closer to  the  interpolation  points have  larger
weights. For instance, when the  field is estimated at the position of
TM$_1$, to  which the magnetometer M$_4$ is  the closest magnetometer,
the value it  measures is the largest contributor  to the interpolated
field in TM$_1$. At the same time, magnetometers M$_1$ and M$_3$ being
nearly  equidistant from both  test masses,  their weights  are almost
identical (see table  \ref{tab:positions} for more details).  Finally,
no  apparent  or  easily  deductible physical  relationship  is  found
between the estimated gradient at the positions of the test masses and
the magnetometer inputs.


\section{Variations of the magnetic dipoles}
\label{chap.3}

The numerical experiments done so far indicate that the neural network
interpolating scheme  offers good performances  when properly trained,
irrespective of its specific architecture.  However, we emphasize that
the neural  network has been  trained using simulated data,  while for
the real spacecraft the neural  network will be trained using onground
measured   data.   This   data,  as   already  mentioned   in  section
\ref{sec:model}, is planned to be obtained in Spacecraft Magnetic Test
Campaign, to be performed at during  2011. To assess how this could be
done we have determined how many  batches of samples need to be fed in
the neural  network to obtain  the desired accuracies.  We  have found
that for  a proper  training of  the network, at  least 10  batches of
samples must be recorded from the real spacecraft with all the sources
of magnetic  field onboard.  Only in  this way we  can be sufficiently
confident on the  trained neural network.  Each of  these batches will
be constituted of  $10^3$ vectors of 28 values  each, corresponding to
12  readings of  the magnetometers  (3 components  for each  of  the 4
magnetometers),  6  magnetic  field  readings  (3  components  of  the
magnetic field  measured at  the positions of  each test mass)  and 10
readings of  the gradients  of the magnetic  field (5 values  for each
test  mass).  This  will allow  to choose  an specific  neural network
model in a realistic case.

\begin{figure}[!t]
\centering
\includegraphics[width=0.9\textwidth]{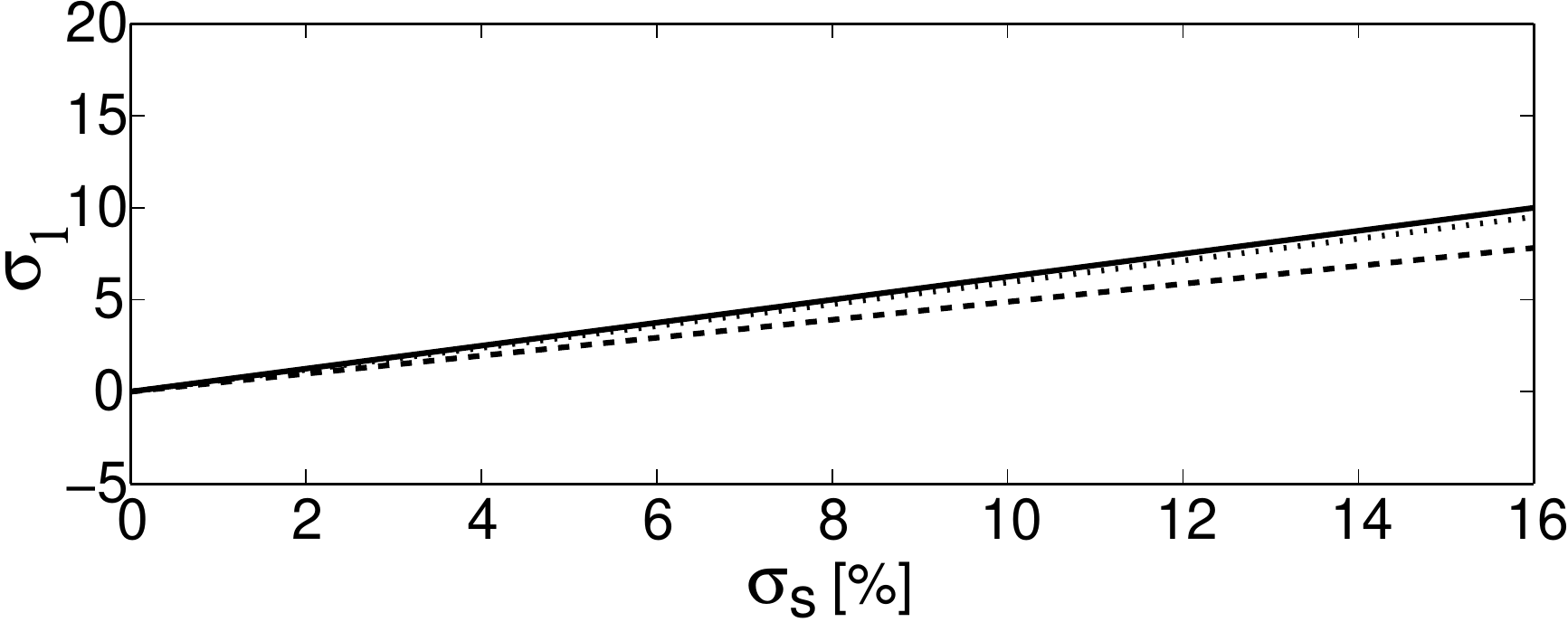}
\includegraphics[width=0.9\textwidth]{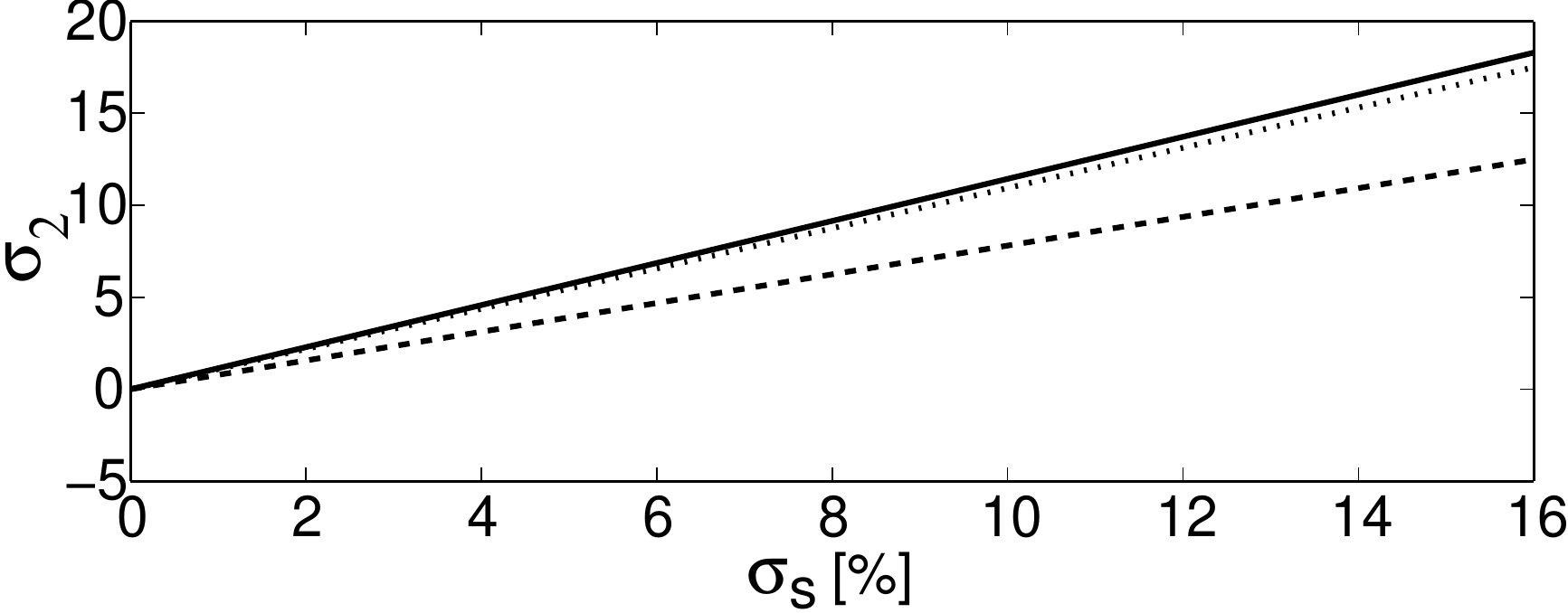}
\caption{Quality   of  the   estimate  (standard   deviation   of  the
         estimation)  as  a function  of  the  real magnetic  inflight
         measurements  with respect to  the onground  measurements (in
         percentage).  Again,  the   solid  line  corresponds  to  the
         $B_x$-component, the dotted  line to the $B_y$-component and,
         the dashed line to the $B_z$-component.}
\label{fig:uncertaintyField}
\end{figure}

It  is expected  that  the  magnetic characteristics  of  each of  the
spacecraft units will not change due to launch stresses.  However, the
measurements taken  onground may not  be accurate enough  to represent
the  real  magnetic  inflight  characteristics of  these  units.   For
instance, some units would  be missing during the onground measurement
campaign  or some  others  can change  their magnetic  characteristics
during the  lifetime of the mission  or, finally, it could  be as well
that the system operation cannot be measured onground accurately.  For
all these reasons the predictions of the neural network may be biased.
Hence, it is important to  assess the robustness of the predictions of
the neural network in front of  changes in the magnetic dipoles of the
electronic   boxes.   To  do   so  we   have  adopted   the  following
procedure. We varied  randomly each of the components  of the magnetic
field  of all  the sources  of  magnetic field  according to  Gaussian
distributions.  The  width of such Gaussians, $\sigma_B$,  is our free
parameter and  corresponds to a  given percentage of deviation  of the
specific component with  respect to that of the  training set. In this
way we can  simulate a difference between flight and  ground data in a
simple and realistic manner.

The   results   obtained   using   this   procedure   are   shown   in
figure~\ref{fig:uncertaintyField},   where   we   show  the   standard
deviation of the probability density function of the estimation of the
three components of the  magnetic field interpolated using the trained
neural network as a function of  the width of the Gaussians. As can be
seen in this figure, the  error of the estimate increases linearly for
increasing widths of the Gaussian.  Nevertheless, our simulations show
that  offsets of $\sim  15\%$ per  component in  each of  the magnetic
sources result  in a global error  of the estimate of  $\sim 15\%$ for
the magnetic  field and  of only  $\sim 5\%$ for  the gradient  at the
positions of  the test  masses, a very  interesting result.   Thus, we
conclude that  our interpolation scheme  is fairly robust in  front of
small differences in the flight-ground data configuration.


\section{Offsets in the magnetometers}
\label{chap.4}

\begin{figure}[!t]
\centering
\includegraphics[width=0.9\textwidth]{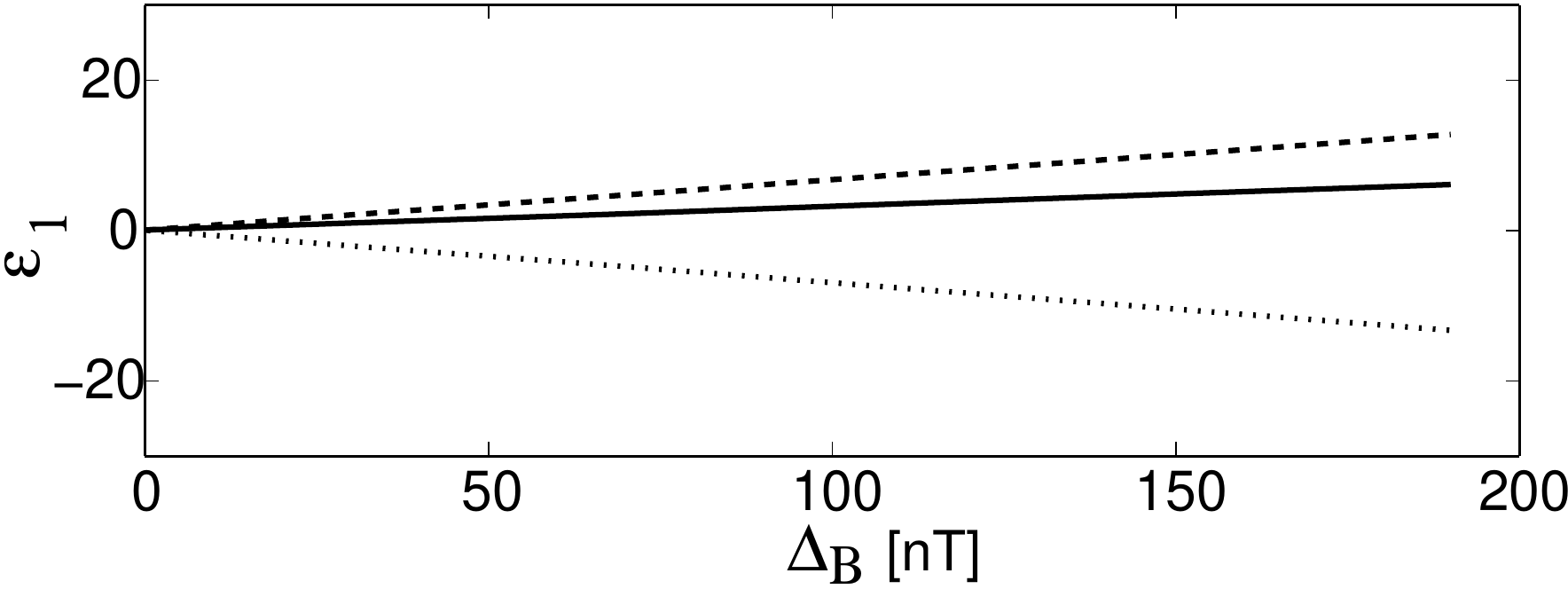}
\includegraphics[width=0.9\textwidth]{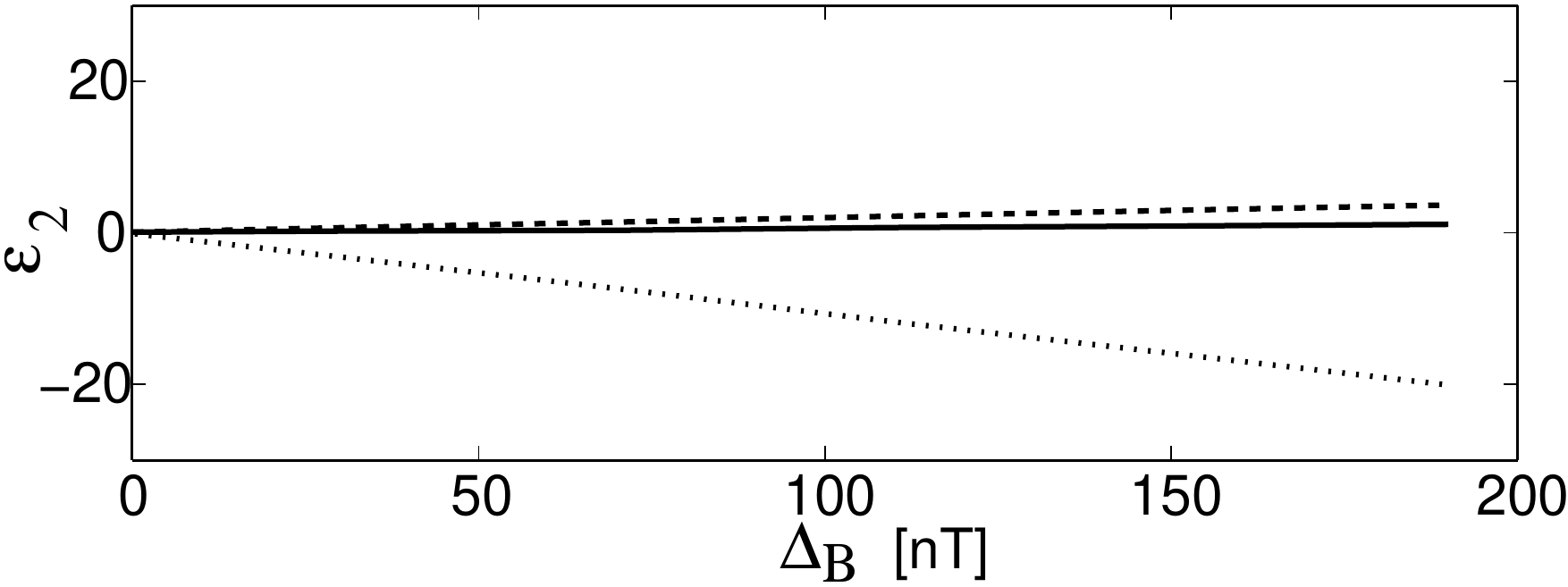}
\caption{Quality of  the estimate  (mean error) as  a function  of the
         magnitude  of the offset  in all  4 magnetometers.  The solid
         line corresponds  to the $B_x$-component, the  dotted line to
         the   $B_y$-component   and,   the   dashed   line   to   the
         $B_z$-component.}
\label{fig:offsetMag}
\end{figure}

It has been shown recently  that the magnetometers readings may suffer
from  an  unpredictable  offsets  \cite{bib:magOffset} due  to  launch
stresses.   In  particular,  this  offset  is most  problably  due  to
temperature changes during launch, and varies from $\Delta_B\sim 1$ nT
to several  nT.  This, of  course, may have important  consequences in
the  estimate  of   the  magnetic  field  at  the   positions  of  the
magnetometers, as  the interpolating algorithm  presented here largely
depends on the reading of the magnetometers.  

To assess the robustness of the interpolation scheme to the offsets in
the readings of the magnetometers we have simulated a random vector of
offsets  (a 12 valued-vector,  1 offset  for each  of the  12 magnetic
channels),   according   to   a   Gaussian   distribution   of   width
$\Delta_B$. This offset vector has been added to the inflight readings
when  performing   the  assessment  of  the  results   output  by  the
interpolation  network.   Several   simulations  have  been  performed
varying $\Delta_B$  from 1  nT to  200 nT.  The  results are  shown in
figure  \ref{fig:offsetMag}.   As  can  be  observed,  the  errors  in
magnetic field estimation are below 10\%  up to an offset level at the
magnetomters of 80 nT --- which  is one order of magnitude larger than
the  offset  observed  in other  space  missions~\cite{bib:magOffset}.
Consequently,  we conclude  that  the magnetic  data  analysis of  the
mission will not be appreciably affected by the possible offset of the
magnetometers readings.


\section{Precision of the position of the magnetometers}
\label{chap.5}

Another aspect which may also be relevant for the determination of the
magnetic field  and gradients at the  positions of the  test masses is
the uncertainty  in the  location of the  heads of  the magnetometers.
Actually, the neural  network is trained with the  nominal position of
the magnetometers, and  the inflight training will be  done with these
nominal  values. The  uncertainty  in these  values  may represent  an
important source of  error because the neural network  learns from the
geometrical distances  between the  test masses and  the magnetometers
--- see table \ref{table:input-ouput}.

\begin{figure}[!t]
\centering
\includegraphics[width=0.9\textwidth]{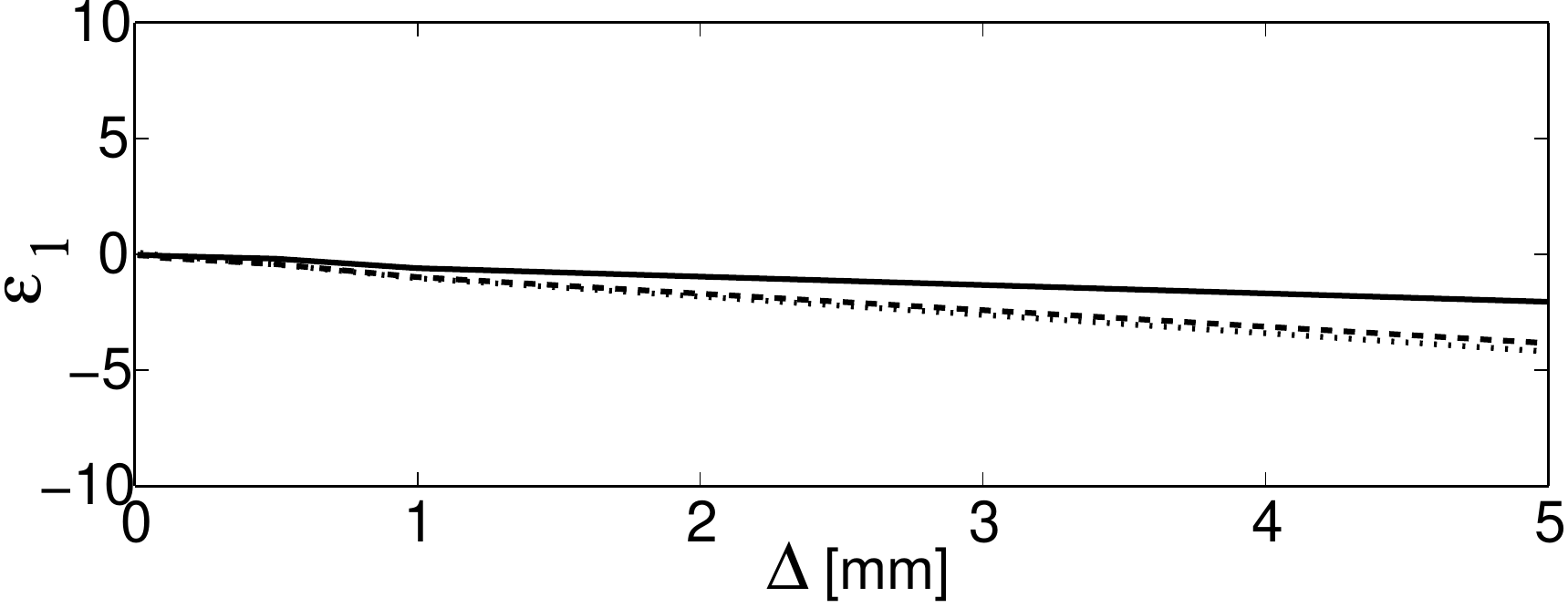}
\includegraphics[width=0.9\textwidth]{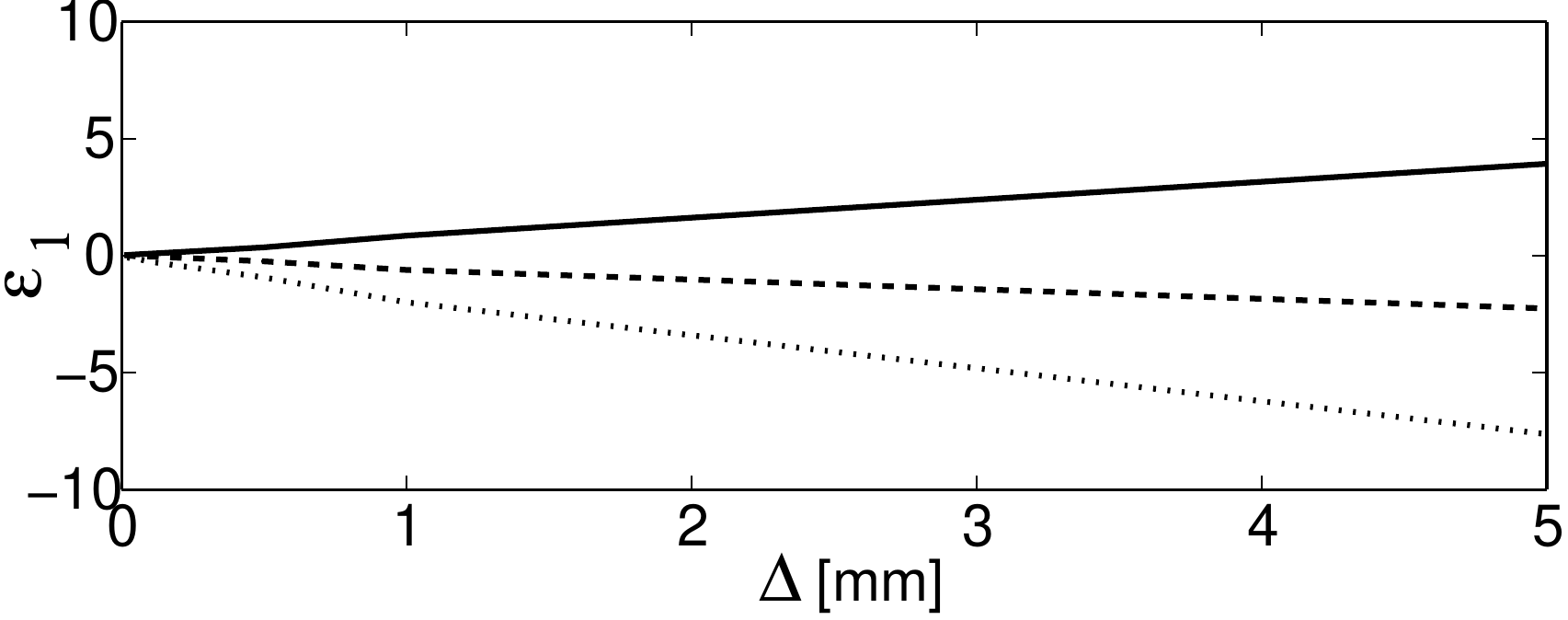}
\caption{Quality of  the estimate (mean  error) due to  the mechanical
         uncertainty in the precise position of the magnetometers.}
\label{fig:posMag}
\end{figure}

The onboard tri-axial magnetometers will be four TFM100G4-S. These are
fluxgate  magnetometers built by  Billingsley. By  construction, these
magnetometers consist  of three different magnetic  sensors, along the
$x$-, $y$- and  $z$-directions.  For each of these  axes, the fluxgate
magnetometer  consists of  a sensing  (secondary) coil  surrounding an
inner drive (primary) coil around permeable core material.  Due to the
large size of  the head of these low-noise  magnetometers, the spatial
resolution in each  of the directions is $\sim 4.0$  mm.  On the other
hand, the coils of the magnetometers have an orthogonality better than
$1^\circ$.   This  angular  error  may  be  transformed  to  a  linear
uncertainty  by  multiplying  by   the  longest  distance  inside  the
magnetometer caging, $l\simeq 82.5$ mm, resulting in an uncertainty of
$\sim 1.5$  mm.  Finally, the  exact placement of the  satellite units
onto the  satellite walls may be  unprecise. It is  estimated that the
mechanical precision will be on the order of the $\mu$m, and therefore
it will be considered negligible in this analysis. The overall spatial
uncertainty  of  the sensing  position  of  the  magnetometers can  be
computed  by adding  in  quadrature the  different contributions,  and
turns out  to be $\Delta\sim 4.3$  mm. In view of  these conundrums we
performed an additional  set of simulations in which  the positions of
the magnetometers where randomly changed within 5 mm. We then computed
the error  in the estimate  of the interpolating neural  network.  The
results  are shown  in  figure \ref{fig:posMag}.  Clearly, the  neural
network  outputs  a mean  error  in  interpolation  below 6\%  if  the
mechanical  uncertainty lies  below 4.3  mm, which  is the  worst case
expected in the mission. Therefore,  the neural network is expected to
be very robust to this kind of uncontrollabale situations.


\section{Varying environmental conditions}
\label{chap.6}

\begin{figure}[!t]
\centering
\includegraphics[width=0.8\textwidth]{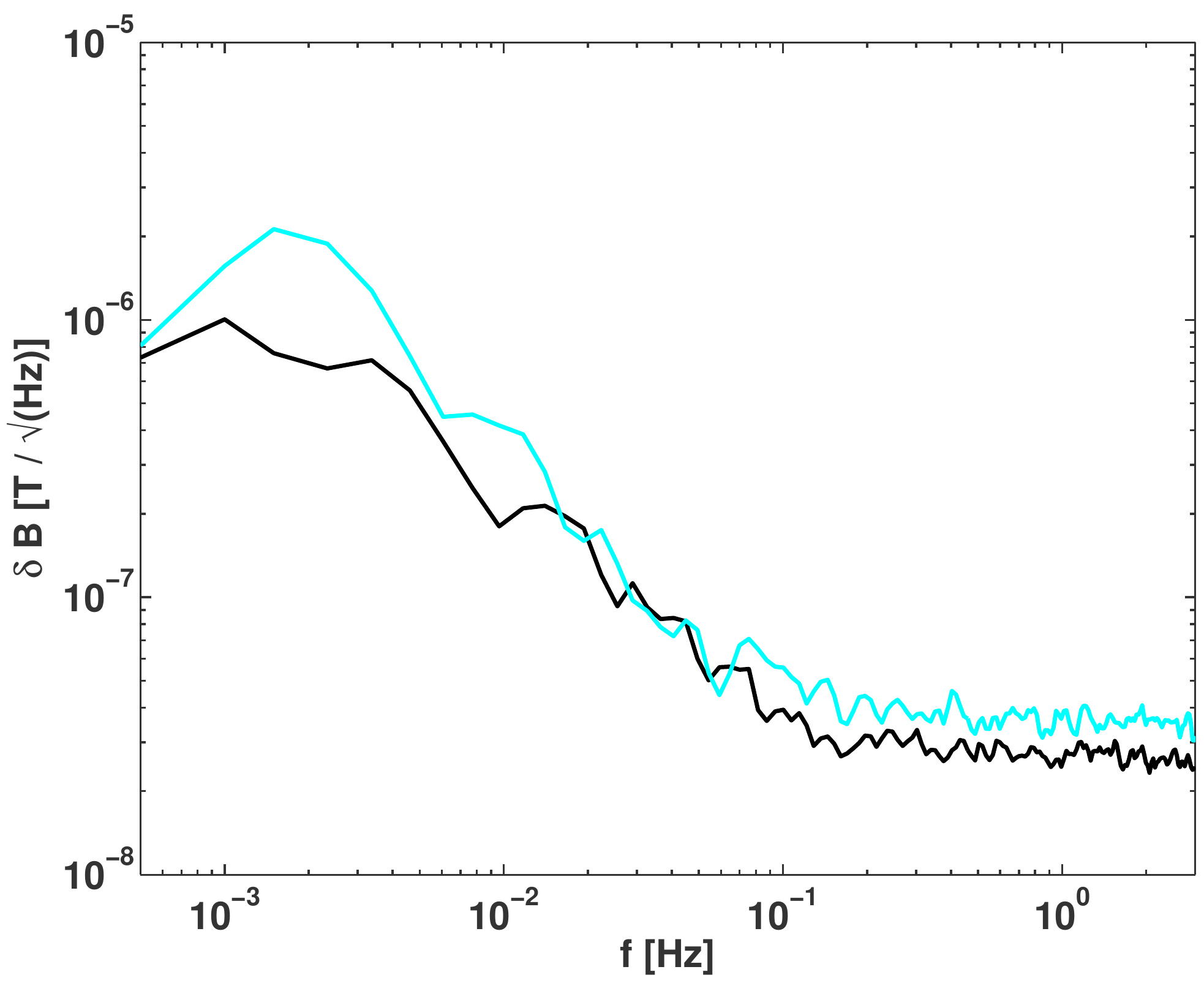}
\caption{Spectrum  of  fluctuations  of  the  magnetic  field  at  the
         position of  the test masses.  The black line  corresponds to
         the spectrum at the position of  test mass 1 and the cyan one
         that at the position of test mass 2.}
\label{fig:spectrumTMs}
\end{figure}

As can be  seen in Eq. (\ref{eq.1}), there  is a non-linear dependence
of the force  on the magnetic field. This  means that the acceleration
depends  on the  temporal variations  of  the magnetic  field and  its
gradient.  Specifically, a coupling of the value of the magnetic field
with the  variations of its  gradient (and viceversa) exists.   In the
previous sections we have  shown that our neural network interpolating
algorithm correctly retrieves the values of the magnetic field and its
gradient at the positions of the  test masses when they are assumed to
do not vary  with time.  However, these quantities  are expected to be
subject to small low-frequency  fluctuations.  Thus, we need to assess
if our method is able to correctly follow a slow drift of the magnetic
field and its gradient.

As  previously mentioned,  the  magnetic  field inside  the  LCA is  a
consequence   of  the   electronic  subsystems   present   inside  the
spacecraft.   Almost all  operational amplifiers  (the  most important
source of noise of the  electronics processing chain of each unit) are
subject to a $1/f$ noise around 0.1 Hz or higher frequencies. Magnetic
tests of  every unit have not  yet been performed, but  it is foreseen
that  the  spectrum of  fluctuations  of  the  magnetic field  at  the
position of the test masses will be very similar to the noise spectrum
of the  amplifiers. In  particular, it is  expected that  the spectrum
will have a  $1/f$ branch below a roll-off frequency of  0.1 Hz, and a
white  noise branch  extending up  to 10  Hz. The  predicted spectrum,
which has  been obtained assuming  a worst-case scenario ---  that is,
assuming an amplitude $5\mu$ A  m$^2/\sqrt{{\rm Hz}}$ at 0.1 Hz --- is
shown in figure \ref{fig:spectrumTMs}.  As it will be shown below, one
of the direct consequences of  including the fluctuations given by the
noise  spectrum of  figure  \ref{fig:spectrumTMs} is  the presence  of
low-frequency  variations  of up  to  300 nT  for  each  of the  three
magnetic components.  These fluctuations may cause important errors in
the  magnetic  field estimation  if  not  considered  in the  training
process. 

\begin{figure}[!t]
\centering
\includegraphics[width=0.8\textwidth]{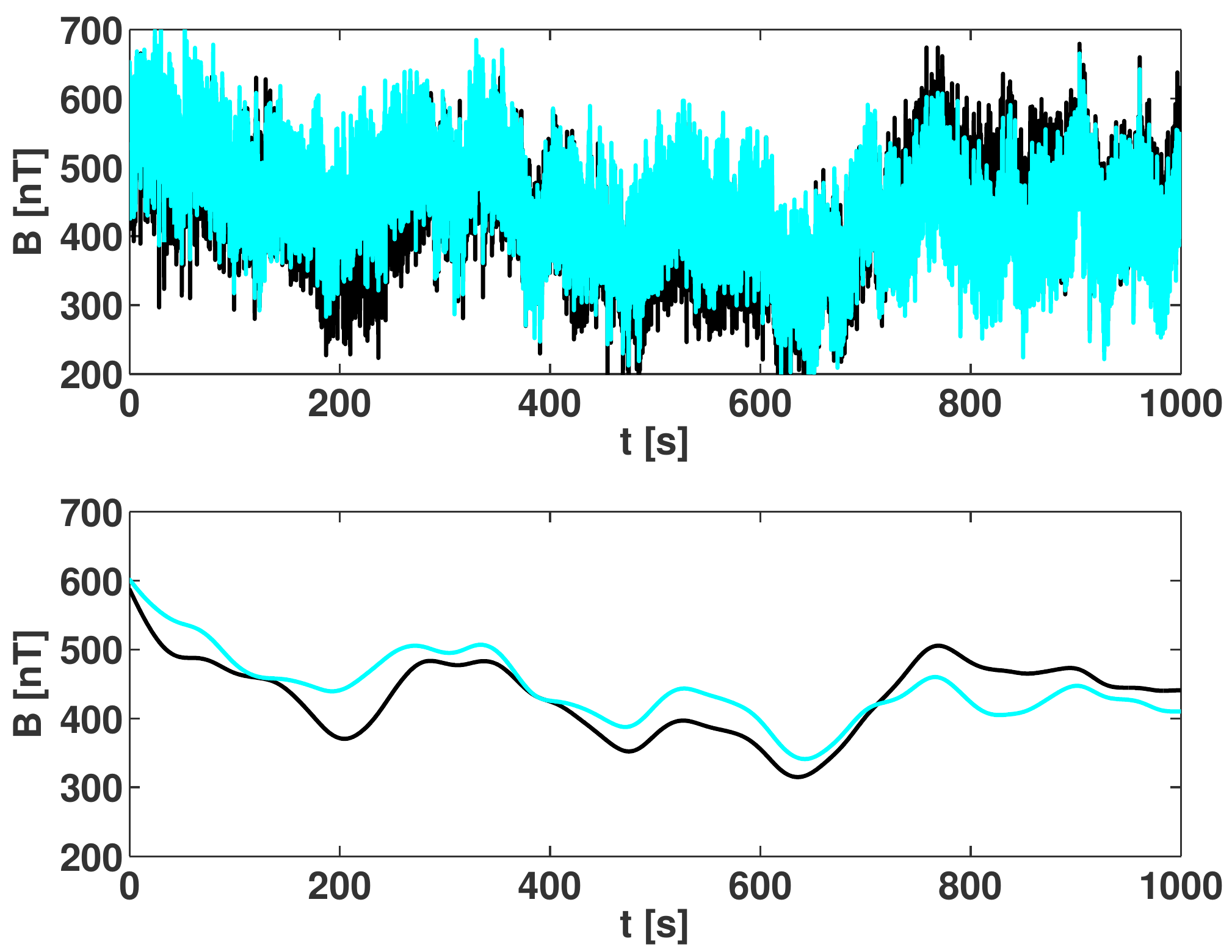}
\caption{Top panel: temporal realization of the magnetic field at test
         mass 1 (black line)  and interpolated magnetic field given by
         the trained  network (cyan line).  Bottom panel: same  as the
         top  panel, but  10 mHz  with a  first order  low-pass filter
         (unitary gain).}
\label{fig:temporal}
\end{figure}

Neural networks can be  classified into dynamic and static categories.
Static networks  have no feedback elements  and, consequently, contain
no delays.   Thus, the  output is calculated  directly from  the input
(and   only  the  current   input)  through   feedforward  connections
\cite{bib:kecman}.  The training of  static networks is performed with
the well  known and efficient backpropagation  algorithm, as described
in section 2.2.   In dynamic networks, the output  depends not only on
the current input to the network,  but also on its previous inputs and
outputs \cite{bib:dreyfus,bib:kecman}.   Thus, for our  case one might
quite naturally  think that  we should be  forced to choose  a dynamic
neural  network.  Nevertheless,  as  shown in  section  2.5, the  most
important feature  of our interpolation  scheme is the ability  of the
neural  network to  learn the  underlying structures  of  the magnetic
field inside the  LCA.  Since training a dynamic  network is hard task
and,  moreover, the learning  rate is  usually very  slow it  is worth
exploring the  possibility of using  instead a static network  with an
adequate training  procedure adapted to  this new scenario.   In other
words, we  have to  let know the  network during the  training process
that  a drift  occurs.   To do  this  we use  a  simple and  effective
training procedure.  We first  generate 10 different time series using
uncorrelated  white noise  realizations. We  then compute  the dynamic
range of the  magnetic field for each of  these realizations. Of these
time  series we  select those  five  which have  the widest  dynamical
range, and we concatenate them.  These  5 time series are then used to
train the network.

\begin{figure}[!t]
\centering
\includegraphics[width=0.8\textwidth]{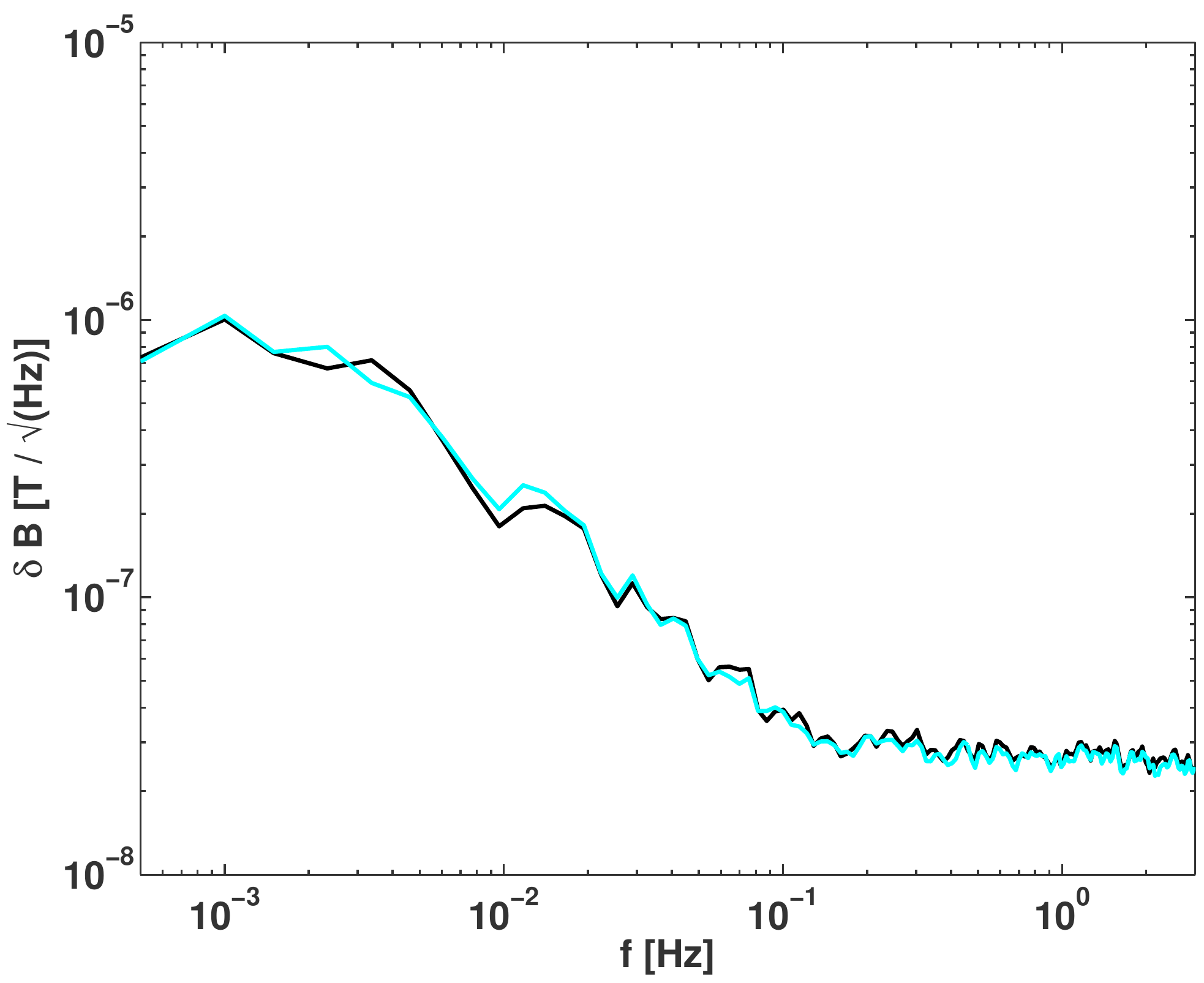}
\caption{Spectral  density of the  magnetic field  at the  position of
         test mass 1 (black line)  compared to the spectral density of
         the  magnetic field  retrieved  by the  neural network  (cyan
         line).}
\label{fig:spectrums}
\end{figure}

\begin{table}
\caption{\label{tab:comparison}Standard deviation  of the error output
         by the network for both the case of a constant magnetic field
         and a fluctuating one.}
\centering
\begin{tabular}{lcccccc}
\hline\noalign{\smallskip}                          
&
\multicolumn{2}{c}{Constant {\bf B}} &
\multicolumn{2}{c}{Fluctuating {\bf B}} \\
\noalign{\smallskip}\cline{2-5}\noalign{\smallskip}
  &  $\sigma_1$ & $\sigma_2$ & $\sigma_1$ & $\sigma_2$ \\
\noalign{\smallskip}\hline\noalign{\smallskip}
 $B_x$    	& 2.68 & 3.28 & 8.75 &  14.75  \\
 $B_y$          & 2.71 & 3.27 & 4.86 &   8.46  \\
 $B_z$ 		& 3.15 & 3.82 & 2.92 &  13.16  \\
 $|B|$         	& 2.13 & 3.15 & 5.85 &  16.30  \\
\noalign{\smallskip}\hline
\end{tabular}
\end{table}

With  this  new  training  technique  the  interpolation  results  are
remarkably  good.  To  illustrate  the goodness  of our  interpolation
procedure in  the top panel  of figure \ref{fig:temporal} we  show the
temporal evolution of  the fluctuating magnetic field for  test mass 1
(black  line) and the  interpolated result  obtained using  the neural
network trained  with the fluctating  examples (cyan line). As  can be
seen,  although  there  are   some  differences,  the  result  of  the
interpolation closely  resembles the actual magnetic  field.  This can
be better appreciated when both signals are filtered at 10 mHz using a
low-pass  filter   (bottom  panel  of  this   figure).   Clearly,  the
interpolated  magnetic field  follows very  closely the  real magnetic
field.  Moreover,  the spectrum of the interpolated  magnetic field is
very similar  to that of  figure \ref{fig:spectrumTMs}. This  is borne
out from figure \ref{fig:spectrums}, in which we compare both spectra.
Clearly, the  interpolated spectrum  (cyan line) follows  very closely
the  real  one  (black  line),  indicating  that  the  neural  network
correctly describes  the physical  properties of the  varying magnetic
field.   The only  remarkable difference  when a  fluctuating magnetic
field is  analyzed is  that in this  case the neural  network performs
slightly worse.  This can be seen in table \ref{tab:comparison}, where
we show the  standard deviations of the estimates  for both a constant
magnetic field and a fluctuating one, for all the field components and
its modulus  and for  both test  mass 1 ($\sigma_1$)  and test  mass 2
($\sigma_2$). As can  be seen, the estimation errors  are larger for a
fluctuating  magnetic field,  as it  should  be expected,  but do  not
increase dramatically.


\section{Summary and conclusions}
\label{chap.7}

LISA Pathfinder  is a  very challenging space  mission, since  it will
test in flight many novel (and critical) technologies which are needed
to  satisfactorily put  in orbit  the first  space-borne gravitational
wave  detector.   More specifically,  this  mission  will measure  and
control  very accurately  the  motion  of two  test  masses in  almost
perfect gravitational free-fall. To  do this, the diagnostic system of
LISA   Pathfinder  will  monitor   with  unprecedented   accuracy  the
disturbances of the  motion of the test masses.   An essential part of
this subsystem is the magnetic diagnostics subsystem, which will be in
charge of  measuring the magnetic  noise. To this end,  this subsystem
has four tri-axial magnetometers,  which due to design constraints are
placed far from the positions  of the test masses. Thus, measuring the
magnetic field  at these positions is  not an easy  task.  To overcome
this problem a  novel approach in which neural  networks were used was
recently  proposed \cite{bib:myPaper}.   The initial  results obtained
using this  technique were  encouraging but a  full assessment  of its
reliability was still lacking. 

Accordingly,  we  have  studied  how different  alternatives  for  the
architecture  of  the  neural  network  affect the  precision  of  the
interpolation of the magnetic field  and its gradients at the position
of  the test  masses.  We  have performed  a study  of  the underlying
structures of the neural network and we have found that the ability of
our interpolating scheme to recover the correct values of the magnetic
field and gradients at the positions  of the test masses is due to the
fact that the neural network is able to learn from the readings of the
magnetometers which  are closest to  the corresponding test  mass, and
that the  most important contribution  for each component  field comes
from the  corresponding magnetomer reading.   We have also  found that
the number of  neurons in the hidden layer  originally proposed is the
optimal one,  and that a larger  number of neurons in  this layer does
not improve  the quality of  the interpolation. Also, the  results are
not sensitive to the choice of the transfer function, and consequently
the  simplest  choice,  a   linear  transfer  function,  is  the  best
option. Finally, we have also  found that the optimal number of hidden
layers is just one. 

We have  also discussed  how the neural  network must be  trained with
real  data. In  particular,  we  stress the  importance  of finding  a
training process  adequate to the  set of data the  magnetometers will
deliver in flight. This underlines  the need to characterize on ground
to our best ability the magnetic field distribution across the LCA for
as many  as possible foreseeable working  conditions. This information
will  be obtained  from the  Magnetic Test  Campaign, to  be performed
during 2011.  Reliable information  on the magnetic characteristics is
essential for  a meaningful assessment  of magnetic noise in  the LTP,
and  may  lead  to  model  various  networks  for  different  magnetic
configurations.  Our results indicate  that when typical variations in
the magnitudes of the magnetic dipoles of the electronic units are fed
into our neural network algorithm  the quality of the estimates of the
magnetic  field and  its  gradients degrade  linearly with  increasing
departures from the onground measurement, although the measurements of
the  magnetic  field  degrade   faster  than  those  of  the  gradient
components.  However,  the quality of  the estimates does  not degrade
dramatically.

We  have also  studied  which would  be  the effect  in the  on-flight
measurements of an  offset in the readings of  magnetometers caused by
temperature changes during launch, and we have found that, for typical
offsets, the  interpolating algorithm works reasonably  well. The same
can be said  about the uncertainty in the position of  the head of the
magnetometers.

Finally,  we have  also assessed  the accuracy  of the  magnetic field
interpolation   when   a   low-frequency   drift   of   the   magnetic
characteristics  is  present,  concluding  that  with  an  appropriate
training procedure, good results  are obtained. Thus, we conclude that
the neural network interpolating  algorithm is robust enough to obtain
a good  estimate of the  magnetic field at  the positions of  the test
masses under most circumstances.


\section*{Acknowledgments}
This work  was partially supported  by MCINN grants  ESP2004-01647 and
AYA08--04211--C02--01.  Part  of this work  was also supported  by the
AGAUR.   We thank  our  anonymous referee,  whose  comments helped  to
improve the manuscript.



\end{document}